\documentclass[aps,pre,twocolumn,superscriptaddress]{revtex4-2}
\usepackage[utf8]{inputenc}
\usepackage{graphicx}
\usepackage{epstopdf}
\usepackage{bm}
\usepackage{amsmath}
\usepackage{verbatim}
\usepackage{amssymb}
\usepackage{listings}
\usepackage{color}
\definecolor{darkblue}{rgb}{0,0,0.6}
\usepackage{here}
\usepackage[colorlinks,linkcolor=darkblue,citecolor=darkblue,urlcolor=darkblue]{hyperref}
\usepackage{mathtools}
\DeclareRobustCommand{\average}[1]{\left\langle #1 \right\rangle}
\renewcommand\vec[1]{\boldsymbol{#1}}

\newcommand\fig[1]{Fig.~\ref{#1}}

\newcommand\eq[1]{Eq.~(\ref{#1})}
\newcommand{\ee}{\text{e}}

\newcommand{\piB}{\pi_\text{\tiny{B}}}

\begin{document}

\title{Irreversible swap algorithms for soft sphere glasses} 

\author{Yoshihiko Nishikawa}

\affiliation{Department of Physics, Kitasato University, Sagamihara, Kanagawa, 252-0373, Japan}

\affiliation{Graduate School of Information Sciences, Tohoku University, Sendai, Miyagi 980-8579, Japan}

\author{Federico Ghimenti}

\affiliation{Laboratoire Mati\`ere et Syst\`emes Complexes (MSC), Universit\'e Paris Cit\'e \& CNRS (UMR 7057), 75013 Paris, France}

\affiliation{Department of Applied Physics, Stanford University, 348 Via Pueblo, Stanford, CA 94305, USA}

\author{Ludovic Berthier}

\affiliation{Gulliver, UMR CNRS 7083, ESPCI Paris, PSL Research University, 75005 Paris, France}

\author{Fr\'ed\'eric van Wijland}

\affiliation{Laboratoire Mati\`ere et Syst\`emes Complexes (MSC), Universit\'e Paris Cit\'e \& CNRS (UMR 7057), 75013 Paris, France}
\affiliation{Yukawa Institute for Theoretical Physics, Kyoto University,
Kitashirakawaoiwake-cho, Sakyo-ku, Kyoto 606-8502, Japan}

\date{\today}

\begin{abstract}
We extend to soft repulsive interaction potentials a recently proposed irreversible swap algorithm originally designed for polydisperse hard spheres. The original algorithm performs rejection-free, irreversible, collective swap moves. We show that event-driven cluster updates of particle diameters can also be performed in continuous potentials by introducing a factorised Metropolis probability. However, the Metropolis factorisation needed to deal with continuous potentials decreases the efficiency of the algorithm and mitigates the benefits of breaking detailed balance. This leads us to propose another irreversible swap algorithm using the standard Metropolis probability that accelerates the relaxation of soft sphere glasses at low temperatures, compared to the original swap algorithm. We apply these efficient swap algorithms to produce very stable inherent structures with vibrational density of states lacking the quasi-localised excitations observed in conventional glasses. 
\end{abstract}

\maketitle

\section{Introduction}

Supercooled liquids experience a dramatic slowing down of their dynamics upon a mild reduction in temperature or increase in density. A fundamental understanding of this phenomenon is the central goal of theories of the glass transition~\cite{berthier2011theoretical}. The slow dynamics of glass-formers poses serious practical challenges for their numerical investigation~\cite{berthier2023modern}. When approaching the glass transition, numerical simulations performed by integrating Newton's equations of motion  experience the glassy slowing down at low temperature, and thermal equilibrium cannot be achieved within practically accessible timescales. Similarly to Newtonian dynamics, Monte Carlo simulations implementing local translational moves reproduce the characteristic dynamics of supercooled liquids on long timescales and experience slow dynamics as well~\cite{Berthier_2007}. However,  Monte Carlo algorithms that sample a given target distribution need not be restricted  to local translations, with irreducibility and aperiodicity being the only additional constraints the Markov chain must abide by~\cite{Levin2017,newman1999monte,krauth2006statistical}. Thus any `unphysical' move, as far as it satisfies these two conditions without affecting the target long-time distribution, can be incorporated into a Monte Carlo algorithm. Devising efficient dynamical rules that reduce the dynamical slowing down extends the range of temperature and densities that can be accessed by numerical simulations, leading to a better understanding of the system properties. Beyond glasses, extended ensemble and cluster algorithms~\cite{Swendsen1987,Wolff1989,Berg1992,Marinari1992,Dress1995,Hukushima1996,santen2000absence,Wang2001} are successful examples of how efficient Monte Carlo algorithms can bypass slow physical pathways, and can bring new insights into the thermodynamics of many complex systems. 

In the physics of glasses and supercooled liquids, clever Monte Carlo algorithms have also been used to compute the nontrivial static properties, including the Franz--Parisi potential and the point-to-set length~\cite{Yan2004,Berthier2013,Jack2016,Berthier2016,Nishikawa2020,guiselin2020random}. The swap Monte Carlo algorithm~\cite{Grigera2001}, simply referred to as `Swap' in the following, which allows for the exchange of particle diameters in polydisperse mixtures, has been particularly impactful on the numerical study of glasses in the last decade. When applied to size polydisperse systems~\cite{berthier2016equilibrium,Ninarello2017,berthier2019efficient,parmar2020ultrastable}, this relatively simple algorithm can accelerate the equilibration of certain models by several orders of magnitude, allowing the simulations to explore equilibrium states even below the experimental glass transition temperature. Starting from stable configurations obtained with the Swap algorithm, simulations employing the physical dynamics have revealed important correlations between the dynamic and static structures of ultrastable glasses~\cite{Guiselin2022,Scalliet2022}. The Swap algorithm was numerically applied to other systems, such as crystals with complex structures and lattice models~\cite{Bommineni2019,Bommineni2020,Nishikawa2020, alfaro2024swap, Nishikawa2024}. However, the Swap itself eventually becomes slow at very low temperatures~\cite{shiraishi2024characterizing}, and the quest for even more efficient Monte Carlo algorithms needs to be continued. 

\begin{figure*}
    \includegraphics[width=\textwidth]{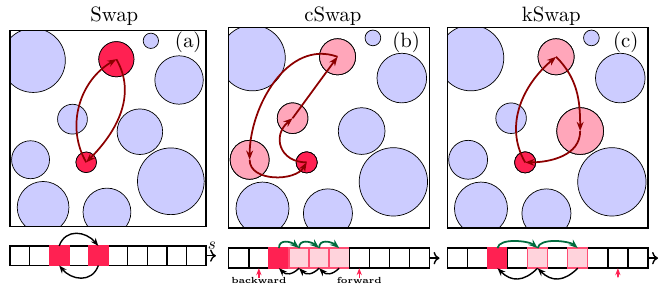}
    \caption{Schematic description of various swap algorithms. Each panel shows a configuration (top) and the one-dimensional array, $s$, listing particle labels in order of increasing diameters. 
    (a) In Swap, two particles are randomly selected along $s$ and a reversible exchange of their diameters is performed using a Metropolis acceptance rule.
    (b) In cSwap, a directed chain of swaps is built starting from an active particle (in bright red). Consecutive particles in $s$ are successively swapped resulting in a collective move. The activity label (red arrow) is updated either to the left (backward cSwap) or to the right (forward cSwap). 
    (c) In kSwap, the consecutive swaps have a fixed size $k \geq 1$ (here $k=2$), and the activity label is updated according to the forward prescription. Typically kSwap clusters contain less particles but span a larger interval in diameter space.}
    \label{fig:swaps}
\end{figure*}

The Swap algorithm samples the Boltzmann distribution through reversible moves, {\it i.e.} that obey detailed balance: it randomly selects two particles and attempts an exchange of their diameters with the standard Metropolis probability. A line of research dating back to the nineties~\cite{hwang1993accelerating, chen1999lifting, diaconis2000analysis, hwang2005accelerating,ghimenti2023sampling} points instead to the fact that giving up on the reversibility of a Markov chain while preserving its stationary distribution, can be rewarded with faster convergence. A most prominent family of algorithms exploiting this idea are the so-called lifted Markov chains~\cite{turitsyn2011irreversible, vucelja2016lifting}. In a nutshell, an additional degree of freedom, the lifting variable, is introduced which drives the system away from equilibrium and breaks detailed balance. The dynamics is nevertheless carefully designed to sample the Boltzmann distribution in its steady state, upon marginalising over the lifting degrees of freedom. This approach was successfully applied in a variety of physical settings~\cite{bernard2009event, bernard2011two, sakai2013dynamics, Michel2014, michel2015event}. Despite the recent progress in the mathematical understanding of lifted Markov chains~\cite{krauth2021event, guyon2023necessary, monemvassitis2023pdmp}, the design of an appropriate scheme for a given physical system remains a nontrivial task. It has recently been shown~\cite{Ghimenti2024irreversible, berthier2024monte} how the lifting approach can be employed to improve the efficiency of Swap for hard particles in two and three dimensions. The introduction of lifted degrees of freedom allows to perform driven, collective updates of the particle diameters. The resulting collective Swap (named `cSwap') algorithm for hard sphere glasses achieves a significant speedup compared to Swap in the dense glassy regime. However, cSwap is tailored for the hard sphere potential and it cannot be applied for continuous potentials. 

Here, we extend the algorithm to soft repulsive potentials, and study its efficiency. We design and compare algorithms that perform driven particle swaps in models of soft repulsive spheres in two and three dimensions. The algorithms studied in this work are illustrated in Fig.~\ref{fig:swaps}. In addition to the standard Swap in Fig.~\ref{fig:swaps}(a), we first discuss a variant of the cSwap algorithm for hard spheres, which differs from the original version in Ref.~\cite{Ghimenti2024irreversible} in the manner the lifting degrees of freedom are updated. We then generalise these two cSwap algorithms for soft sphere continuous potentials, which exploit the factorised Metropolis probability~\cite{Michel2014} to perform directed collective moves, see Fig.~\ref{fig:swaps}(b). Our simulations show that, contrary to hard spheres, the cSwap algorithms become less efficient than the reversible Swap algorithm for polydisperse soft sphere models. We prove that the probability factorisation is the cause of this somewhat disappointing result. This leads us to design another irreversible swap algorithm, which we name `kSwap'. This algorithm employs the full Metropolis probability and performs consecutive directed swap of a fixed jump size $k$, see Fig.~\ref{fig:swaps}(c). We find that kSwap is more efficient than Swap and cSwap for these continuous potentials. As an application, we use these new algorithms to produce inherent glass states with original physical properties.  

The manuscript is organised as follows. We present the Swap and cSwap algorithms, and their extensions to continuous potentials in Sec.~\ref{sec:collective swap}. The models used to benchmark these algorithms are introduced in Sec.~\ref{sec:models}. The efficiency of these algorithms is studied in Sec.~\ref{sec:time}. The new kSwap algorithm is introduced and studied in Sec.~\ref{sec:new}. Inherent glassy states are produced and analysed in Sec.~\ref{sec:inherent}, while Sec.~\ref{sec:conclusion} concludes the manuscript.

\section{Collective swap algorithms}

\label{sec:collective swap}

\subsection{Revisiting the hard sphere system}

\label{sec:cSwap_hard}

The collective swap (cSwap) algorithm~\cite{Ghimenti2024irreversible,berthier2024monte} performs driven cluster updates in the space of particle diameters. This is done by extending the configuration space of the system through an extra degree of freedom. The system then evolves  through a driven Markovian dynamics in the extended phase space. The dynamical rules are designed in such a way that the nonequilibrium stationary state is the Boltzmann distribution. This general construction can be rewarded with a faster convergence toward the target stationary distribution than equilibrium dynamics, and it is known in the literature as a lifted Markov chain~\cite{chen1999lifting,diaconis2000analysis,Bernard2009, Michel2014, michel2015event, nishikawa2015event, krauth2021event}. 

In the collective swap algorithm the particle labels are sorted in a one-dimensional array by increasing diameter. An activity label, $v$, is initialised at random among the $N$ possible particle labels. This activity label $v$ constitutes the additional lifting degree of freedom. The active particle serves to initiate the collective swap move. An exchange of diameters between particle $v$ and the particle $i$ immediately to its right in $s$ is proposed, where $s$ is the one-dimensional array that lists particle labels in order of increasing diameters. This exchange is accepted with a probability given by the Metropolis rule. For hard spheres, the move is accepted with probability $1$ if the exchange of diameters generates no overlaps, and rejected otherwise. If the move is accepted, the diameters of $v$ and $i$ are exchanged and the array is updated. A new attempt at updating the diameter of $v$ with the nearest right neighbour is again performed, following the same procedure. The accumulation of a number $n_c$ of accepted moves of this type is equivalent to an overall permutation where particle $v$ is maximally inflated, while all other particles involved in the move are simultaneously deflated and replace their left neighbour in the array $s$. 

The inflation of the active particle $v$ stops when an exchange in the chain produces an overlap. When this happens, the lifting degree of freedom $v$ is updated instead of simply rejecting the exchange, thus terminating a single collective Swap move. In the original version of the algorithm~\cite{Ghimenti2024irreversible,berthier2024monte}, the particle immediately to the left of the cluster becomes active for the next step, see Fig.~\ref{fig:swaps}(b), thus moving backward along the diameter array. We refer to this choice as the backward collective Swap. It is also possible to choose the particle immediately to the right of the updated cluster, see Fig.~\ref{fig:swaps}(b). In this case, the activity label moves in the same forward direction as the inflating particle along the sorted array. We christen this algorithm the forward collective Swap.

In a simulation, sets of cSwap moves alternate with sets of translational Metropolis Monte Carlo moves. A set of $N_\text{cSwap}$ cSwap moves is performed with probability $p_\text{cSwap}$, and a  sweep made by $N$ translational moves is performed otherwise. Here $N_\text{cSwap}$ is a number comparable with the number of particles in the system, $N_\text{cSwap} \sim N$. Finally we observe that in both forward and backward versions of cSwap, the activity label $v$ is updated deterministically. One way to warrant ergodicity is to intersperse this deterministic evolution with a resetting process: After a cluster update, the activity label $v$ is resampled uniformly along the array $s$ with a small probability $p_\text{r}$. The precise value of $p_\text{r}$ does not impact the efficiency of the algorithm, as long as $p_\text{r}\sim N^{-1}$. Here we take  $p_\text{r} = \frac{1}{N_\text{cSwap}}$.  

The computational cost of the cluster update can be noticeably reduced by resorting to an event-driven approach. Instead of attempting the pairwise swaps of the active particle $v$ one by one, the largest particle $i^*$ with which $v$ can be exchanged without generating overlaps is easily identified through a search operation, {\it e.g.} binary search \cite{Knuth1998art3}, along $s$. The identification of particle $i^*$ defines the entire cluster to be updated in the collective Swap move, namely particles between $v$ and $i^*$. The cluster move is then performed by updating the diameter of $v$ to the one of $i^*$, and by updating the diameter of all the other particles in the cluster, $i^*$ included, to the diameter of the particle on their left along the sorted array. The resulting cascade of deflation can be safely performed, as the shrinking of a hard sphere never generates overlaps with its surrounding neighbours. This event-driven approach makes the computational cost of a collective update comparable with the one of a single equilibrium swap move in Swap, thus allowing cSwap to maintain an edge over equilibrium Swap in practical implementations.

\subsection{Generalisation to continuous potentials}

The cSwap algorithm described above needs reconsideration when continuous potentials are used, as finite energy differences now need to be handled correctly. In soft spheres, Monte Carlo moves which produce particle overlaps involving finite positive increments in energy have a finite probability to occur. Given an assignment of the particle diameters, the probability to accept an exchange of the diameters of two particles $i$ and $j$ in a system at inverse temperature $\beta$, is given by the Metropolis probability
\begin{equation}\label{eq:Metropolis}
    \begin{split}
        P_\text{Met} &= \min[ 1, \ee^{-\beta \Delta E(i,j|s)}] = \ee^{-\beta \max[0, \Delta E(i,j|s)]}\,,
    \end{split}
\end{equation}
where $\Delta E(i,j|s)$ is the change of energy of the system via the proposed swap move. Cluster moves can of course be constructed out of this Metropolis acceptance probability, but the evaluation of Eq.~\eqref{eq:Metropolis} would become numerically very expensive when the size of the cluster becomes large as $\Delta E$ would now involve a large number of energy evaluations. As observed in Ref.~\cite{michel2015event}, this problem can be alleviated by replacing the acceptance probability in Eq.~\eqref{eq:Metropolis} with a factorised Metropolis probability $P_\text{fMet}$, defined as
\begin{equation}
    P_\text{fMet}(i, j | s) = \ee^{-\beta \max[0, \Delta E_{ij}(s)]}\times \ee^{-\beta \max[0, \Delta E_{ji}(s)]},
    \label{eq:factorised Met}
\end{equation}
where $\Delta E_{ij}(s)$ is the energy difference of particle $i$, and of particle $i$ only, before and after the exchange of its diameter with particle $j$ on the sorted array $s$. The total energy change in the swap move is thus $\Delta E(i,j|s) = \Delta E_{ij}(s) + \Delta E_{ji}(s)$. The key observation is that $P_\text{fMet}$ also obeys detailed balance and can be used to sample the Boltzmann distribution. For repulsive soft interactions, which we use below, the shrinking particle has a negative energy change while the inflating one has a positive change. If the diameter of particle $i$ inflates, the factorised Metropolis probability in Eq.~(\ref{eq:factorised Met}) reduces to
\begin{align}
    P_\text{fMet}(i, j | s) = \ee^{-\beta \Delta E_{ij}(s)}\,.
\end{align}

Using the factorised Metropolis probability, we propose a generalisation to soft repulsive potential of the forward cSwap algorithm that is amenable to an event-driven formulation. As in the hard sphere case, the activity label is initialised at random along the sorted array $s$. Particle $v$ attempts to exchanging its diameter with the particle on its right along $s$. We denote this particle by $v_1$. The exchange is accepted according to the factorised Metropolis probability $P_\text{fMet}(v,v_1|s)$. If accepted, the diameters and the array $s$ are updated, after which $v$ attempts a new exchange. This procedure is iterated until the first rejection. The activity label is then updated according to the forward cSwap rule. As for hard spheres, a uniform resampling of the activity label is implemented to ensure aperiodicity. This procedure defines the forward cSwap algorithm for a soft repulsive potential.  

The steady state of the forward cSwap algorithm is given by the Boltzmann distribution, since the latter verifies the stationarity, or global balance, condition
\begin{equation}\label{eq:global_balance}
    \pi_\text{\tiny{ss}}(s|v) = \sum_{s',v'} \phi\left((s',v') \to (s,v)\right)\,.
\end{equation}
Here $\pi_{\text{\tiny{ss}}}(s|v)$ is the stationary probability distribution for an instance of ordered diameters $s$ and of the lifted label $v$, defined as
\begin{equation}\label{eq:steady_state}
    \pi_{\text{\tiny{ss}}}(s|v) = \frac{1}{N}\piB(s)\,,
\end{equation}
with $\piB(s) \propto \ee^{-\beta E(s)}$ the Boltzmann distribution. The quantity $\phi\left((s',v') \to (s,v)\right)$ is the steady state probability flux from a configuration $(s',v')$ to the configuration $(s,v)$. The forward cSwap rules imply that
\begin{equation}
    \begin{split}
        \sum_{s',v'} \phi\left((s',v') \to (s,v)\right) &=\pi_\text{\tiny{ss}}(s^*|v)\ee^{-\beta \Delta E_{v,v_1}(s^*)} \\
        &+ \pi_\text{\tiny{ss}}(s|v_{-1})(1-\ee^{-\beta\Delta E_{v_{-1},v}(s)})\\
        &= \frac{1}{N}\piB(s) = \pi_\text{\tiny{ss}}(s|v)\,. \label{eq:forward_balance_condition}
    \end{split}    
\end{equation}
The configuration $(s^*,v)$ is the configuration that flows into $(s,v)$ upon a successful exchange of particle $v$ with its right neighbour along $s$. The configuration $(s,v_{-1})$ flows into $(s,v)$ after a rejection of the exchange of the particle diameters $v_{-1}$ with $v$. Here, and in what follows, $v_n$ denotes the particle $n$ steps away from $v$ along the ordered array $s$ (if $n$ is positive $v_n$ lies to the right of $v$, if $n$ is negative $v_n$ lies to the left of $v$). The second equality follows from the definition of the steady state and Boltzmann distributions in Eq.~\eqref{eq:steady_state} and the fact that 
\begin{equation}
   \piB(s^*) \ee^{-\beta \Delta E_{v,v_1}(s^*)} = \piB(s)\ee^{-\beta\Delta E_{v_{-1},v}(s)}\,.
\end{equation} 
By plugging Eq.~(\ref{eq:forward_balance_condition}) into Eq.~\eqref{eq:global_balance} we confirm that the stationarity condition is indeed satisfied.

The use of the factorised Metropolis filter allows us to to easily calculate the probability to accept $n$ successive swaps along the array $s$ as 
\begin{align}
    \prod_{i=v}^{v_n} P_\text{fMet}(i, i_1 | s) = \ee^{-\beta \Delta E_{v,v_n}(s)}\,.
\end{align}
The simplification of the Boltzmann factors in this product is the main rationale behind the introduction of the factorised Metropolis filter. As for hard particles, we can find $i^* = v_n $ efficiently in an event-driven way. To this end, we draw a uniform random number $\Upsilon \in (0, 1)$, and find $i^*$ such that 
\begin{align}
    \ee^{-\beta \Delta E_{v,i^*_{-1}}(s)} < \Upsilon < \ee^{-\beta \Delta E_{v,i^*}(s)}.
\end{align}
Finding $i^*$ can be efficiently done with computational complexity $O(\log N)$ through algorithms such as binary search. A cluster move of all the particles between $v$ and $i^*$ on the array $s$ can then be performed. 

An event-driven backward cSwap algorithm for soft potentials is constructed analogously, by changing the update rule for the lifting degree of freedom. A proof of stationarity for the event-driven backward cSwap is obtained as follows. Given a configuration $(s|v)$, we denote by $m$ the integer such that $v_m$ is the rightmost label along the ordered array $s$. In the backward cSwap algorithm, there are $m$ probability flows into state $(s|v)$. The right hand side of the stationarity condition, Eq.~\eqref{eq:global_balance}, becomes
\begin{align}
    \begin{split}
        \pi_{\tiny{\text{ss}}}(s |v) + \sum_{n=1}^{m-1} \left[\phi_\text{in}(n+1) - \phi_\text{out}(n)\right]\,,
    \end{split}
\end{align}
where 
\begin{align}
    \phi_\text{out}(n) &= \pi_{\tiny{\text{ss}}}(s_n | v_n)\ee^{-\beta \Delta E_{v_1, v_{n+1}}(s_n)},\\
    \phi_\text{in}(n) &= \pi_{\tiny{\text{ss}}}(s_n | v_n ) \ee^{-\beta \Delta E_{v_1, v_n}(s_n)}\,,
\end{align}
and $s_n$ is the configuration that flows into $s$ after performing an event-driven backward Swap move from the active particle $v_1$. Since
\begin{align}
        E(s_n) + \Delta E_{v_1,v_{n+1}}(s_n) = E(s_{n+1}) + \Delta E_{v_1,v_{n+1}}(s_n),
\end{align}
we have $\phi_\text{in}(n+1) - \phi_\text{out}(n) = 0$ for any $n$, and the total probability flow is $\pi(s | v_1) = \pi(s | v)$, which completes the proof. 

Note that the backward algorithm balances the probability flows of multiple cluster moves, not of single exchanges. Thus, interrupting a cluster update violates the balance condition, whereas it does not cause any problem in the forward algorithm, as the latter satisfies the balance condition through every single diameter exchange during the construction of the cluster.

\section{Glass-forming models used for benchmarking}

\label{sec:models}

To test these new cSwap algorithms, we explore their efficiency in the context of two models of glass-formers. We use polydisperse soft spheres in two~\cite{berthier2019zero} and three dimensions~\cite{Ninarello2017} with number density $\phi = 1$, interacting through the following potential: 
\begin{align}
    &V(r_{ij}) / v_0 = 
    \left(\frac{d_{ij}}{r_{ij}} \right)^{12} + c_0 + c_2 \left(\frac{r_{ij}}{d_{ij}} \right)^2 + c_4 \left(\frac{r_{ij}}{d_{ij}} \right)^4,
\end{align}
if $r_{ij} < r^{\text{cutoff}}_{ij}$ with $r^{\text{cutoff}}_{ij} = 1.25d_{ij}$ and $v_0 > 0$ an energy scale, and $V(r_{ij}) = 0$ otherwise. Here, $r_{ij} = |\vec r_i - \vec r_j|$ with $\vec r_i$ the position of particle $i$. The particle diameter is non-additive, $d_{ij} = \frac{d_i + d_j}{2} \left(1 - \epsilon|d_i - d_j|\right)$ with $\epsilon = 0.2$, which prevents crystallisation and demixing. We use the following power-law distribution for the diameters: 
\begin{align}
    P(d) = \left\{ \begin{array}{cl}
    A d^{-3} & : d_\text{min} < d < d_\text{max},\\
    0 & :\text{otherwise},
    \end{array}\right.
\end{align}
where $d_\text{min} / d_\text{max} = 0.45$. The constant factor $A$ is a normalisation constant. This distribution leads to an overall polydispersity $\delta = \left(\overline{d^2}-\overline{d}^2\right)^{1/2} / \overline{d} = 0.23$. The units of length and temperature are $\overline d$ and $v_0$.

The particle positions are updated with the standard Metropolis translational dynamics, where we first displace a particle by a random vector $\vec \delta \in (-\delta_\text{max}, \delta_\text{max})^d$ with $d$ the spatial dimension, and then accept it with the Metropolis probability. We set $\delta_\text{max} = 0.175\overline d$ and $\delta_\text{max} = 0.1\overline d$ for the two- and three-dimensional models, respectively. For the swap algorithms, we perform a set of diameter swaps with probability $p_\text{swap} = 0.2$ and otherwise perform a set of $N$ translational moves. The value of $p_\text{swap}$ is approximately optimised to minimise the relaxation time at low temperatures. When using the event-driven schemes for the collective algorithms, $N_\text{cSwap} = 512$ clusters are updated per unit time, while for the swap algorithms using the full Metropolis probability, one unit time consists of $N$ swap attempts. We verified that this choice of time units faithfully mirrors, up to an overall rescaling, the CPU time required by each algorithm to run. The system size is $N=1024$ throughout the paper, unless explicitly mentioned otherwise.

For the two-dimensional model, we measure the time correlation of the local orientational order \cite{flenner2015fundamental} at time $t$,
\begin{align}
    &C_{6}(t) = \frac{\average{\sum_j \psi_j(0) \psi^*_j(t)}}{\average{\sum_j |\psi_j|^2}},\\
    \label{eq:psi_time_corr}
    &\psi_j(t) = \frac1{|\partial j(t)|} \sum_{k \in \partial j(t)} e^{-6i\theta_{jk}(t)},
\end{align}
where $\theta_{jk}(t)$ is the angle of $\vec r_k(t) - \vec r_j(t)$ with respect to the $x$ axis, and $\average{\cdot}$ represents an average over initial equilibrium configurations. The set of neighbouring particles $\partial j(t)$ for particle $j$ is found using the Voronoi tessellation for each configuration at time $t$. On the other hand, the relaxation of the three-dimensional model is studied using the overlap function
\begin{align}
    Q(t) = \average{\frac1N \sum_i w(|\vec r_i(0) - \vec r_i(t)|)},
\end{align}
where $w(r)$ is $0$ when $r > 0.2 \overline{d}$ and $1$ otherwise. 

In all cases, we define the relaxation time $\tau_\alpha$ as the time where the time correlation functions decrease to the value $\ee^{-1}$. While this is a measurement of a specific correlation time, its temperature dependence faithfully reflects the evolution of the equilibration timescale that characterises either the approach to stationarity or the ergodic exploration of the configuration space at thermal equilibrium. 

\section{Equilibration times of collective swap algorithms}

\label{sec:time}

\begin{figure}[t]    
    \includegraphics[width=\linewidth]{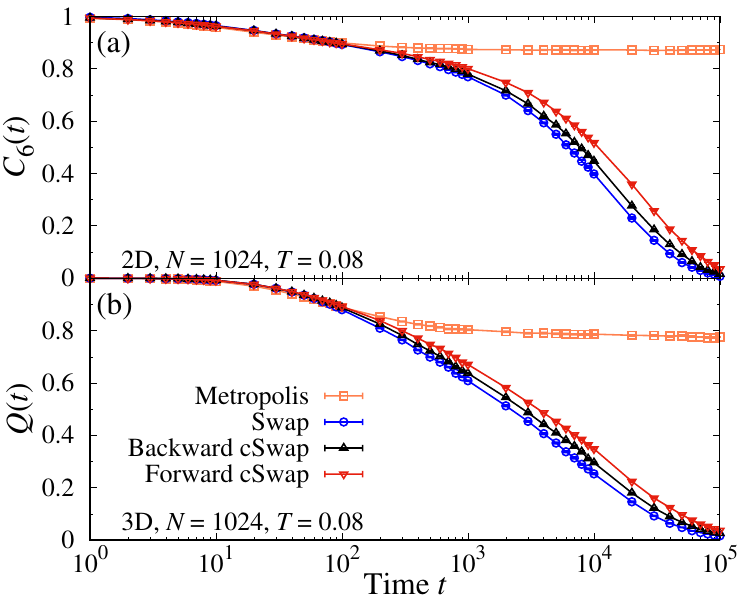}
    \caption{Time correlations of the two dimensional (a) and three-dimensional (b) models at $T = 0.08$ comparing Metropolis translations, with Swap, forward and backward cSwap algorithms. In both cases, Swap remains marginally faster than cSwap algorithms.}
    \label{fig:Time correlation}
\end{figure}

We test the two cSwap algorithms on the models at $T = 0.08$, much lower than the onset temperatures $T_o \approx 0.3$ in two dimensions and $T_o \approx 0.2$ in three dimensions, below which the heterogeneous glassy dynamics sets in \cite{Sastry1998,keys2011excitations}. The data in \fig{fig:Time correlation} show the time correlations for the simple Metropolis, Swap and both cSwap algorithms. At this temperature, the Metropolis translational dynamics is completely frozen and the corresponding time correlation shows a plateau covering the longest simulated  timescale, showing that relaxation is impossible within our simulation window using simple Monte Carlo simulations. On the other hand, all three swap algorithms can decorrelate the particle positions or relative orientations, and all time correlations decay to a small value after $t \approx 10^5$. Strikingly, however, the two cSwap algorithms are both slightly slower than the simple Swap despite the large number of clusters updated per unit time. This result is in stark contrast with the hard sphere situation~\cite{Ghimenti2024irreversible,berthier2024monte}, where cSwap is roughly $10$ times faster than Swap in both spatial dimensions. 

The main difference between the two families of algorithms is the use of the factorised Metropolis probability for the continuous potentials, in order to satisfy the stationarity condition. We may therefore expect the loss of efficiency of cSwap for soft spheres to be caused by the factorisation. For hard spheres, both expressions for the Metropolis probabilities become identical, whereas for soft spheres we generally have $P_\text{fMet} < P_\text{Met}$.  

\begin{figure}
    \centering
    \includegraphics[width=\linewidth]{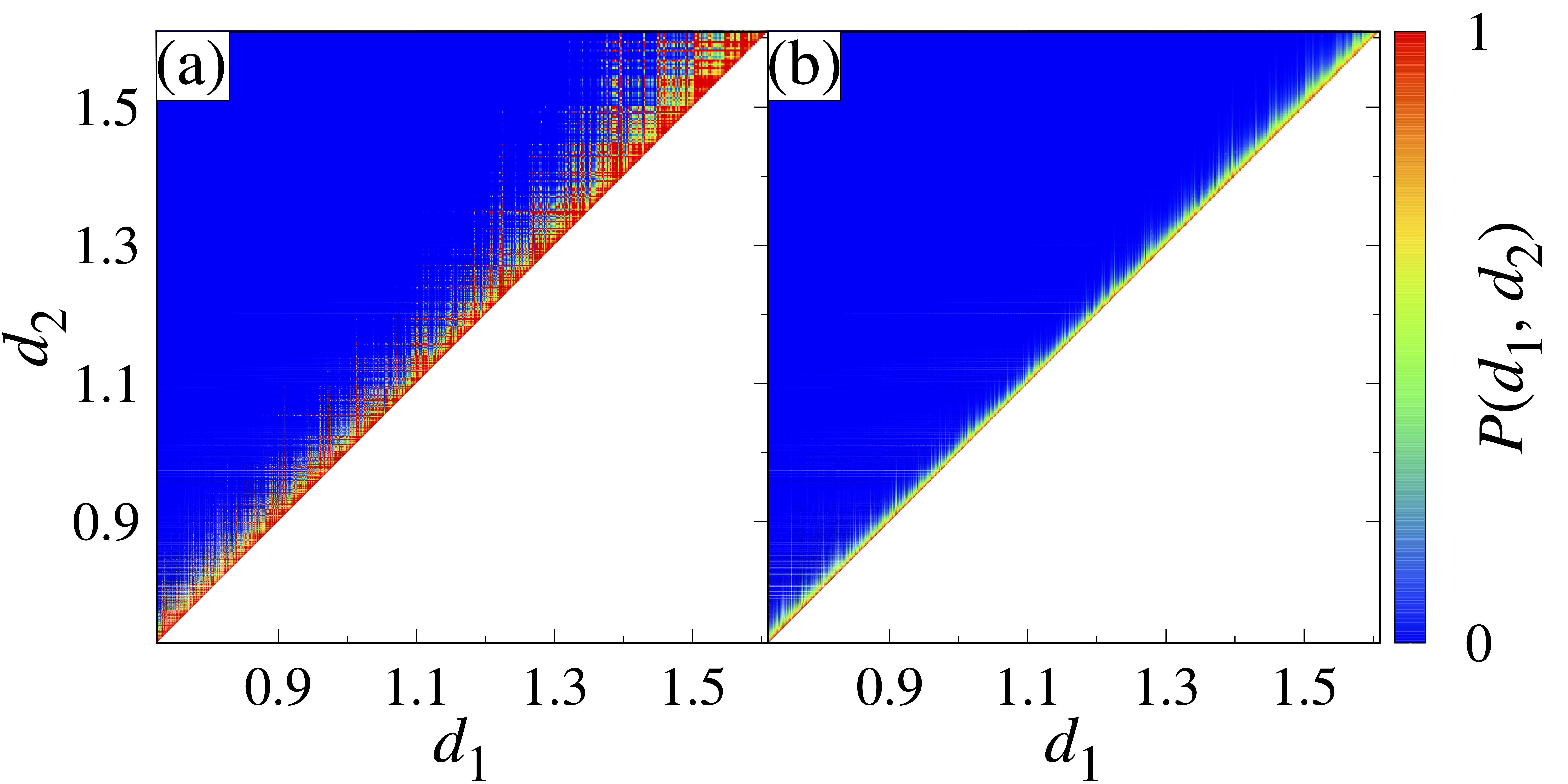}
    \caption{Swap probability $P(d_1, d_2)$ between particles with diameters $d_1$ and $d_2$ at equilibrium in the three-dimensional model at $T = 0.08$, using (a) the full Metropolis expression in \eq{eq:Metropolis} and (b) the factorised Metropolis probability, \eq{eq:factorised Met}, respectively. Off-diagonal elements corresponding to large diameter exchanges are strongly suppressed by the factorised probability in (b).}
    \label{fig:swap prob}
\end{figure}
 
We test this hypothesis in \fig{fig:swap prob}, which compares full and factorised swap acceptance probabilities. It can be seen that the factorisation significantly reduces the acceptance of swap moves at low temperatures. The full Metropolis probability can be close to $1$ even when the two diameters are dissimilar, as the negative and positive energy changes can cancel each other. By contrast, the factorised probability only uses the positive energy change of the inflating particle and rapidly decreases with the diameter difference.

Since the cSwap algorithms gradually increase the diameters of particles by successive moves between neighbouring particles in the $s$ array, the factorised probability results in clusters with a relatively small number of particles ($\approx 20$ at the lowest temperature), leading to slow decorrelation in  diameter space. In fact, the factorised Metropolis probability also slows down the event-chain algorithm for translations for continuous potentials compared with hard particles~\cite{Bernard2009, Michel2014}. We have explicitly checked the influence of the factorised probability in our algorithm by comparing the equilibrium Swap against a version of forward cSwap that uses the full Metropolis probability, with the same number of cluster updates per unit time as in the factorised case. This unfactorised algorithm is faster than Swap, thus demonstrating that the factorised probability is indeed responsible for the efficiency loss. However, a forward cSwap with a large number of cluster updates using the full Metropolis acceptance probability is not practically useful, because of the large overhead needed to estimate $P_\text{Met}$.

\section{A new irreversible swap algorithm with large jumps} 

\label{sec:new}

The generalisation of cSwap to continuous potentials involves a factorisation of the Metropolis acceptance rule. When applied to large clusters, the factorisation decreases the acceptance so much that the algorithm becomes inefficient. Yet the efficiency of cSwap over Swap relies on the rapid inflation of the active particle to the maximum value allowed by the Boltzmann distribution. Large changes of the particle are indeed the source of the speedup offered by swap moves~\cite{Ninarello2017}.

These arguments and the search for an efficient algorithm suggest to introduce collective and directed swap moves that use the Metropolis probability and allow for large moves in diameter space. To this end, we generalise the forward cSwap so that it now incorporates larger jumps, of size $k$ along the $s$ array, see Fig.~\ref{fig:swaps}(c). We christen this algorithm `kSwap'. kSwap is a modification of cSwap where an active particle $v$ is swapped with another particle $v_k$, $k \geq 1$, further away along the one-dimensional array $s$ with the full Metropolis probability (recall that cSwap swaps adjacent particles $v$ and $v_1$ with the factorised Metropolis probability). The jump size $k$ is sampled from a uniform distribution between $1$ and $k_\text{max}$ every time we start the swap update, while it is kept fixed until the update stops. The parameter $k_\text{max}$ thus controls the typical jump size. We incorporate these large jumps into the forward algorithm. Since it uses the full Metropolis probability, the unit time of kSwap consists of $N$ swap attempts. Its computational complexity is thus equivalent to that of Swap.

\begin{figure}[t]
    \includegraphics[width=\linewidth]{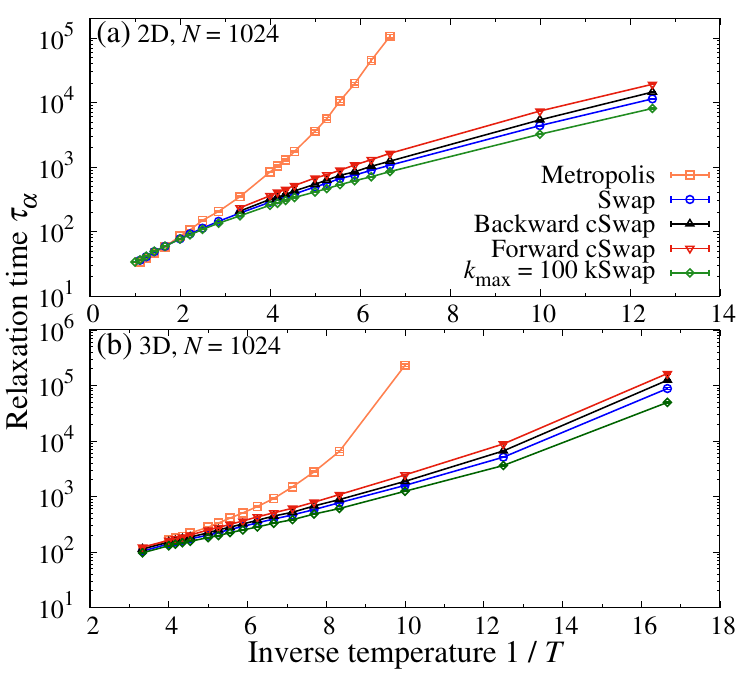}
    \caption{Relaxation time $\tau_\alpha$ as a function of inverse temperature for (a) the two- and (b) three-dimensional systems. The backward and forward cSwap algorithms are slower than Swap, whereas kSwap with $k_\text{max}=100$ is faster.}
    \label{fig: relaxation time}
\end{figure}

Large jumps in diameter space typically yield a smaller acceptance probability, and kSwap is not as collective as the cSwap algorithms, with the typical cluster size $\average{n_c} \approx 3$ when $k_\text{max} = 100$, while $\average{n_c} \approx 20$ for the cSwap algorithms. Nevertheless, kSwap is faster than cSwap and even faster than the original Swap. In \fig{fig: relaxation time}, we show the relaxation time $\tau_\alpha$ of the two  models as a function of inverse temperature $1/T$, for all algorithms. All algorithms incorporating swap moves are much faster than the original Metropolis Monte Carlo. For soft spheres, kSwap is faster than Swap, which is itself slightly faster than both forward and backward cSwap. The speedup of kSwap with respect to Swap weakly depends on temperature and increases towards low temperatures. 

The relaxation time of kSwap depends on $k_\text{max}$. For $k_\text{max} = 1$, the algorithm reduces to a forward cSwap with the full Metropolis probability, and is slower than Swap. We find, for $N=1024$ and $k_\text{max} \in [50, 150]$, that kSwap is faster than Swap, that the value $k_\text{max}=100$ is roughly optimal and that it yields the shortest relaxation time at low temperatures. For continuously polydisperse models, the typical interval between two neighbouring diameters in the array decays as $N^{-1}$, thus we expect $k_\text{max}$ to linearly scale with $N$ for optimal performance. We have checked $k_\text{max} = 50$ indeed yields almost the same relaxation time for kSwap when $N = 512$, supporting the above scaling of $k_\text{max}$ with $N$.

\section{Using swap algorithms to produce non-conventional stable glasses}

\label{sec:inherent}

Algorithms such as gradient descent and FIRE (for Fast Inertial Relaxation Engine)~\cite{Bitzek2006} starting from an equilibrium configuration at temperature $T$ produce an energy minimum, also called inherent structure (IS). Inherent structures and the related concept of a potential energy landscape are frequently used to discuss the properties of supercooled liquids and glasses~\cite{goldstein1969viscous,Stillinger1982,Stillinger1988,Stillinger1995,Kurchan1996,Buchner1999,Sciortino1999,Sciortino2005,Heuer2008,khomenko2020depletion,Nishikawa2022}. Two quantities are of particular interest. First, the value of the energy density in the inherent state reached from a given $T$ reveals the stability of the corresponding amorphous structures. Typically, an IS has a lower energy when it is reached from a lower $T$, indicating more stable glassy structures lying deeper in the potential energy landscape. The second quantity is the vibrational density of states (DOS), which describes the range of eigenfrequencies characterising eigenmodes of the vibrational matrix at the energy minimum. Roughly speaking, an abundance of low frequency modes would correspond to an energy minimum with many soft vibrational modes. Remarkably, the vibrational density of states $D(\omega)$ of inherent glassy states displays a power law scaling at small frequencies, $D(\omega) \sim \omega^4$~\cite{Lerner2016,Mizuno2017,Kapteijns2018,Wang2019}, in addition to the Debye law. The stability of the glass is usually encoded into the prefactor of this power law, which decreases sharply with increasing stability~\cite{Wang2019}. The $\omega^4$ scaling of the DOS is thus a universal property of amorphous solids, which suggests an abundance of low-frequency soft modes resulting from the disordered structure of the IS. 

We generate inherent structures from equilibrium configurations at temperature $T_\text{init}$ using zero-temperature Monte Carlo algorithms, in which only updates lowering the total energy are accepted. For the Metropolis translational dynamics, we reduce the jump size to $\delta_\text{max} = 0.0005\overline{d}$. We refer to this as the Metropolis quench. We also use hybrid Monte Carlo quenches using both translational and swap moves, using either Swap or kSwap. Note that cSwap algorithms do not work for quenches, as the factorised probability is always zero for collective swaps at zero temperature. The number of Monte Carlo sweeps per particle in our simulations ranges from $10^6$ to $2\times 10^6$. The number of negative eigenvalues of the Hessian matrix for a generated configuration after the quench is typically zero, indicating that the Monte Carlo approach indeed produces energy minima.

\begin{figure}[t]
    \centering
    \includegraphics[width=\linewidth]{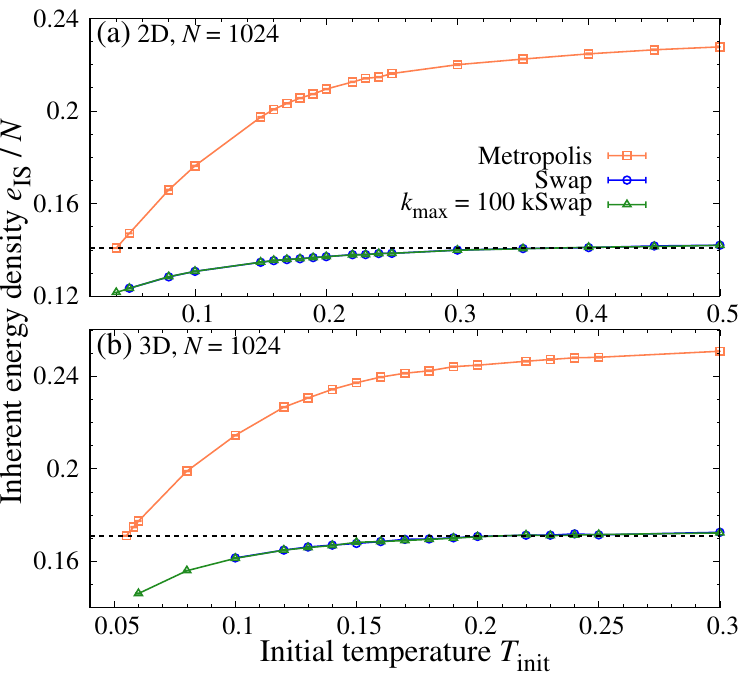}
    \caption{Inherent structure energy $e_\text{IS} / N$ as a function of initial temperature $T_\text{init}$ for (a) the two- and (b) the three-dimensional systems, respectively. The dashed line in each panel represents $e_\text{IS} / N$ at the lowest $T_\text{init}$ for the Metropolis quench.}
    \label{fig:IS energy}
\end{figure}

In \fig{fig:IS energy}, we show the IS energy density $e_\text{IS} / N$ as a function of initial temperature $T_\text{init}$ for the Metropolis, Swap, and $k_\text{max}=100$ kSwap algorithms. The Metropolis quench reproduces the behaviour observed previously in glass-formers, with a very weak dependence of the IS energy at high $T_\text{init}$, and a sharper dependence below a crossover temperature, corresponding to the onset temperature~\cite{Sastry1998}. Below the onset, IS energies depend more sensitively on the initial temperature, lower $T_\text{init}$ ending deeper in the potential energy landscape. The crossover correspond roughly to the onset of slow dynamics, near $T_\text{init} \approx 0.3$ in two dimensions and $T_\text{init} \approx 0.2$ in three dimensions~\cite{Guiselin2022}.

A clear observation in Fig.~\ref{fig:IS energy} is that quenching from the same configuration using a zero-temperature algorithm that performs both translational and swap moves yields much lower energies than Metropolis quenches. The performance of the two swap algorithms are nearly equivalent. The temperature dependence of the IS energies is less pronounced than for conventional quenches, expecially at high temperatures. and it is essentially the same for the two swap algorithm. There is no obvious qualitative difference between two and three dimensions. In both cases, we observe that instantaneous quenches starting from very high temperatures yield inherent states that are extremely deep in the potential energy landscape, and in fact much deeper than what could be obtained without swap algorithms.  

These findings  for continuous potentials confirm the results obtained for fast compressions of hard particles with the swap Monte Carlo algorithm~\cite{berthier2016equilibrium,Ghimenti2024irreversible,berthier2024monte}, and rationalise the efficiency of approaches where gradient descent involving not just translational degrees of freedom are used to produce stable glassy structures as developed in \cite{brito2018theory,kapteijns2019fast,hagh2022transient,corwinref}. Qualitatively, using swap moves during annealing seems much more efficient than very slowly cooling the system without swap moves. We note in Fig.~\ref{fig:IS energy}(a) a qualitative difference between hard and soft potentials. For two dimensional hard disks, fast compressions with swap moves seemingly all terminate at the same density, independently of the initial state~\cite{Ghimenti2024irreversible}. This observation led Bolton-Lum {\it et al.} to conclude that this end-point represents an `ideal' disk packing~\cite{corwinref}. For continuous potentials, we do not find evidence for such an `ideal' amorphous structure (or ground state) that would be systematically found in the potential energy landscape.  

\begin{figure}[t]
    \centering
    \includegraphics[width=\linewidth]{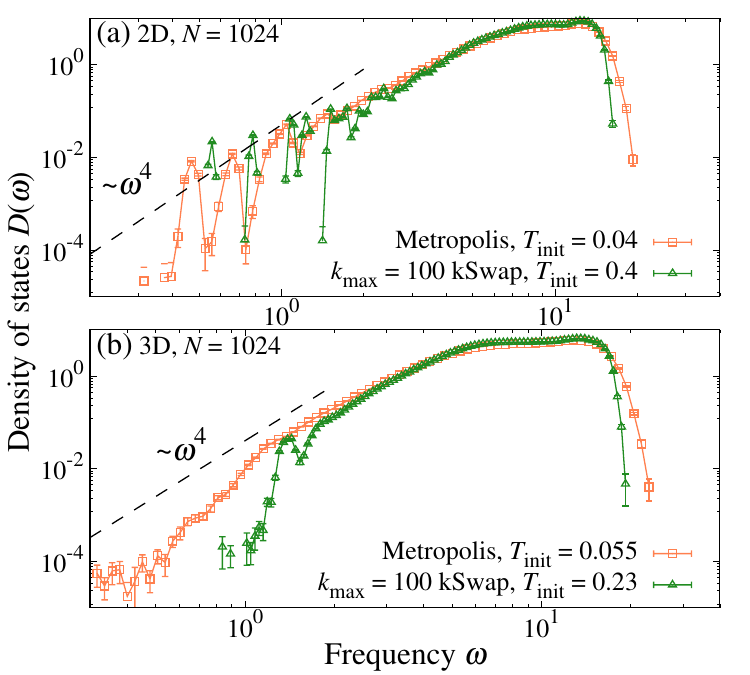}
    \caption{Vibrational density of states of inherent structures for (a) two and (b) three dimensions. Temperatures are chosen so that IS energies are the same in the two algorithms, see dashed lines in Fig.~\ref{fig:IS energy}. The swap moves deplete the soft modes at low frequencies.}
    \label{fig:DOS}
\end{figure}

Using swap moves to prepare inherent structures not only produces much lower energies, but also affects the physical properties of the energy minima that are found. This is revealed by the vibrational DOS. In~\fig{fig:DOS}, we show the DOS for two and three dimensional glasses and compare in each case inherent structures that have the exact same energy (dashed lines in Fig.~\ref{fig:IS energy}) but are produced either with or without swap moves. Clearly, the DOSs are very different for the two quench protocols, showing that the IS energy does not uniquely characterise amorphous structures. For the two-dimensional system, Fig.~\ref{fig:DOS}(a), the Metropolis quench has several isolated modes extending towards low frequencies with an envelope compatible with the $\omega^4$ power law reported before. These few low-frequency modes are nearly removed by the swap quenches. This effect becomes more obvious in three dimensions, Fig.~\ref{fig:DOS}(b), where the smooth tail found for Metropolis quenches is replaced by a gapped distribution. This again shows that swap moves during the approach to the energy minimum remove the soft vibrational modes and change the nature of the energy minima that are reached. The observation of gapped vibrational spectra was also reported using augmented gradient descent techniques~\cite{kapteijns2019fast,hagh2022transient}.  

\section{Conclusion}

\label{sec:conclusion}

This paper extends to soft continuous potentials our recent efforts~\cite{Ghimenti2024irreversible,berthier2024monte} to develop irreversible versions of the swap Monte Carlo algorithm to speedup even further the equilibration of glass-forming models. The broader rationale is the belief that any algorithm developed for hard spheres should have an efficient counterpart for continuous potentials, as illustrated before by the event-chain and geometric cluster algorithms, and even the event-driven molecular dynamics~\cite{Alder1957,Dress1995,Liu2004,Bernard2009,Peters2012,Michel2014,Bouchard-Cote2018,rapaport2004art,krauth2006statistical}. 

Here we successfully generalised the irreversible cSwap algorithms developed for hard spheres to soft potentials, following the path used for event-chain translational moves. However, the performance of the resulting algorithms applied to soft sphere models are not as good as expected. We have identified the factorisation of the Metropolis probability as the root cause of the degradation of the algorithmic performance, which led us to devise an irreversible swap algorithm that uses the full Metropolis probability and borrows from both cSwap and Swap. This new algorithm accelerates equilibrium relaxation at low temperature compared to Swap and cSwap, despite the fact that it is less collective than cSwap. Our results provide instructive insights into the nature of collective moves needed in Monte Carlo algorithms for dense glassy systems, and the edge that can be achieved by tuning the range of the irreversible moves. Future research directions should aim at improving further Monte Carlo algorithms, for instance by exploiting, among others, reinforcement learning~\cite{galliano2024policy} and normalising flows strategies~\cite{jung2024normalizing}.

\acknowledgments
We thank Manon Michel and Ernest van Wijland for useful discussions. Y.N. acknowledges support from JSPS KAKENHI (Grant No.~22K13968). F.G., L.B. and F.v.W. acknowledge the financial support of the ANR THEMA AAPG2020 grant.

\bibliography{refs}

\end{document}